\documentclass[conference]{IEEEtran}
\IEEEoverridecommandlockouts

\usepackage{cite}
\usepackage{amsmath,amssymb,amsfonts}
\usepackage{algpseudocode}
\usepackage{graphicx}
\usepackage{textcomp}
\usepackage{xcolor}
\usepackage{braket}
\usepackage{breqn}
\usepackage{algorithm}
\usepackage{multirow}

\def\BibTeX{{\rm B\kern-.05em{\sc i\kern-.025em b}\kern-.08em
    T\kern-.1667em\lower.7ex\hbox{E}\kern-.125emX}}
\begin{document}

\title{Statistical Quantum Phase Estimation: Extensions and Practical Considerations\\
}

\author{\IEEEauthorblockN{ Amit Surana}
\IEEEauthorblockA{
\textit{RTX Technology Research Center}\\
East Hartford, USA \\
amit.surana@rtx.com}
\and
\IEEEauthorblockN{Brandon Allen}
\IEEEauthorblockA{
\textit{RTX Technology Research Center}\\
East Hartford, USA  \\
brandon.allen@rtx.com}
}

\maketitle

\begin{abstract}
We present several refinements and extensions of the statistical quantum phase estimation (SQPE) framework to address some of its key practical limitations, improving its applicability to realistic cases. Recently, a family of statistical approaches for QPE have been proposed where each run uses only a few ancillae and shorter circuits than standard QPE and thus is better suited for early fault-tolerant quantum computers that are qubit-and depth-limited. SQPE method within that family estimates the cumulative distribution function (CDF) associated with spectral density of the Hamiltonian for a given trial state by using its Fourier approximation and then identifies the first jump discontinuity of the CDF to determine the ground state energy (GSE) of the Hamiltonian. It relies on random compilation procedure based on linear combination of unitaries (LCU) decomposition of the Hamiltonian assuming positive Pauli weights and requires a good estimate of lower bound on the overlap between the trial and true ground state, both of which may be difficult to achieve in practice. We address these limitations by generalizing the random compilation procedure for negative Pauli weights and employing a changepoint detection method for determining GSE which does not rely on an estimate of this overlap. We also show that by exploiting symmetry of the Fourier series one can reduce number of circuit runs/samples by a factor of 2x while keeping the GSE estimation accuracy the same. We illustrate these new developments numerically via a quantum simulator in Qiskit. 
\end{abstract}


\section{Introduction}

Quantum computers can be used to simulate dynamics and learn the spectra of quantum systems, such as interacting particles comprising complex molecules or materials, described by some Hamiltonian $H$. Quantum Phase Estimation (QPE) \cite{kitaev1995quantum,nielsen2010quantum} on the unitary $U = e^{iHt}$ efficiently solves the spectral problem of computing ground state energy (GSE) of $H$, whenever one can efficiently prepare a trial state with non-trivial (not exponentially small) overlap with the ground state. The cost of QPE typically depends on the Hamiltonian sparsity $L$, the number of terms in its linear combination of unitary (LCU) decomposition in a suitable basis, such as the Pauli basis. For electronic structure problems in chemistry and materials science, typically $L$ scales as $L=O(N^4)$ for an $N$-orbital problem. Trotter based approaches \cite{poulin2014trotter,babbush2015chemical} while conceptually simple to implement with minimal use of ancilla qubits, have $O(L)$ gate complexity which can be prohibitive for such applications. Qubitization-based implementations \cite{babbush2019quantum,berry2019qubitization} on other hand achieve sub-linear non-Clifford complexity of $O(\sqrt{L}+N)$ but require $O(\sqrt{L})$ ancillae, which increases the qubit cost from $O(N)$ to $O(N^2)$. Recently, statistical approaches for QPE have been proposed \cite{o2019quantum,dutkiewicz2021heisenberg,lin2022heisenberg,spe} where each run uses only a few ancillae and shorter circuits than standard QPE and thus is better suited for early fault-tolerant quantum computers (EFTQC) that are qubit- and depth-limited. In these approaches, a single run generates a sample of an estimator for $U^j$ for some runtime $j$. Multiple runs are repeated with randomly sampled $j$ and statistical analysis of resulting data reveals spectral information with a confidence that increases with the amount of data collected. 

In this paper we will focus on statistical quantum phase estimation (SQPE) approach proposed in \cite{spe} which is doubly randomized in that it not only randomly samples $j$ but then approximates $U^j$ using a random gate sequence based on a random compilation lemma leveraging the LCU decomposition of $H$. Specifically, SQPE estimates the cumulative distribution function (CDF) associated with spectral density of $H$ for a given trial state and then identifies the first jump discontinuity of the CDF to determine the GSE. For this, CDF is approximated as a Fourier series and the random compilation lemma is applied to further decompose $U^j$ in the series expansion into unitaries formed from multi-qubit Pauli gates and Pauli rotation operators. Given this representation of CDF, low-depth Hadamard test circuits are randomly sampled and the measured outputs are combined to obtain an unbiased estimate of CDF.  Given a lower bound on the overlap $\eta$ between the trial and true ground state, a binary search is wrapped over this CDF estimation procedure to determine the GSE within prescribed tolerance with high probability. Unlike other related approaches, in SQPE all approximation and compilation errors can be expressed in terms of statistical noise that is suppressed by collecting more data samples. This allows for a trade-off between the gate complexity per circuit run/sample and the number of samples. Furthermore, this trade-off curve can be numerically computed by formulating a tractable optimization problem and selecting algorithmic parameters which minimize the gate/sample complexity \cite{spe}. 

While the SQPE is a promising framework for EFTQC, there are several practical limitations of the approach: i) the random compilation lemma assumes non-negative Pauli weights in the LCU  decomposition which may not hold true for most quantum chemistry problems of interest;  and ii) estimate of $\eta$ is in general difficult to obtain which can significantly impact SQPE performance and efficiency. In this work we develop several refinements and extensions of the SQPE framework to address some of these practical challenges, improving its efficiency and applicability to realistic cases. The main contributions of the paper are:
\begin{itemize}
  \item We extend the random compilation lemma to deal with negative Pauli weights in the LCU decomposition of given Hamiltonian, making SQPE applicable in a general setting for GSE estimation.
  \item We apply a changepoint detection method for identifying GSE which does not rely on knowledge of $\eta$.
  \item We show that by exploiting symmetry of the Fourier series representation of CDF, one can reduce number of circuit runs/samples by a factor of 2x while keeping the GSE estimation accuracy the same.  
\end{itemize}
The paper is organized as follows. In Section~\ref{sec:speframe} we summarize the SQPE mathematical setting and derive the generalized randomized compilation lemma and ACDF computation leveraging symmetry in Fourier representation. In Section~\ref{sec:spealgo} we outline the modified SQPE algorithm along with its complexity analysis. Other practical considerations are discussed in Section~\ref{sec:prac}. Simulation results are presented in Section~\ref{sec:results} and the paper is concluded with directions for future research in Section~\ref{sec:conc}.

\section{Statistical Quantum Phase Estimation Framework} \label{sec:speframe}

\subsection{Preliminaries}
We will denote by $\mathbb{R}$ as the set of real numbers, $\mathbb{C}$ as the set of complex numbers, $\mathbb{N}$ as the set of natural numbers, $\mathbb{Z}$ as the set of integers, $i$ is imaginary unit, and $\overline{z}$ as a complex conjugate for a complex number $z\in \mathbb{C}$. We will denote the sign function as $\text{sgn}$. We will use standard braket notation in representing the quantum states. 
%small bold letters as vectors, capital bold letters as matrices/operators,  $A^*$ as the vector/matrix complex conjugate, $A^T$ as the vector/matrix transpose and $I_s$ as an Identity matrix of size $s \times s$.

\subsection{Setting}
Let $H$ be a Hamiltonian of interest with $\beta_k$ and $\ket{\psi_k}$ for $k=0,\cdots,K-1$ being the eigenvalues and eigenstates of $H$, respectively. Hence
\begin{equation*}
H=\sum_{k=0}^{K-1}\beta_k \Pi_k, 
\end{equation*}
where $\Pi_k=\ket{\psi_k}\bra{\psi_k}$ is projection onto $k$-th eigenspace of $H$. Let $H$ be decomposed into a linear combination of $n$-qubit Pauli operators $P_i$, i.e., 
\begin{equation*}
H=\sum_{j=1}^{L} \alpha_j P_j, \quad \lambda=\sum_{j=1}^{L}|\alpha_j|,
\end{equation*}
where $\alpha_j$ are in general real-valued scalars. We normalize $H$ with $\tau$ such that
\begin{equation*}
\tau \|H\|\leq \tau\lambda<\frac{\pi}{2},
\end{equation*}
which for instance can be achieved by choosing
\begin{equation*}
\tau=\frac{\pi}{2\lambda+\Delta}<\frac{\pi}{2\lambda},
\end{equation*}
where $\Delta>0$.

For a given trial state $\rho=\ket{\phi}\bra{\phi}$, let $p_k=\text{Tr}(\rho\Pi_k)=|\braket{\psi_k|\phi}|^2$ be the square of the overlap of $\ket{\phi}$ with eigenvectors of $H$. Define a spectral measure of $\tau H$ associated with $\rho$
\begin{equation}
p(x)=\sum_{k=0}^{K-1}p_k\delta(x-\tau\beta_k),\quad x\in[-\pi,\pi],\label{sec:pdf}
\end{equation}
where note eigenvalues of $\tau H$  are $\tau\beta_k$. Since
\begin{equation*}
\tau|\beta_k|\leq \tau\lambda<\frac{\pi}{2}, 
\end{equation*}
$p(x)$ is supported in the interval $(-\pi/2,\pi/2)$.  Extend $p(x)$ to a periodic function with period $2\pi$, i.e. $p(x)=p(x+2\pi)$.

Following~\cite{spe} we wish to calculate a cumulative distribution function (CDF) associated with $p(x)$, defined as
\begin{equation}
C(x)=\text{Tr}(\rho \Pi_{\leq x/\tau}).\label{eq:cdf1}
\end{equation}
If $C(x)$ can be constructed, then eigenvalues of $H$ can be identified through its jump discontinuities. 
%Using standard definition, let $C(x)$ be the CDF associated with the PDF $p(x)$
%\begin{equation*}
%C(x)=\text{Tr}(\rho \Pi_{\leq x/\tau}).
%\end{equation*}
To facilitate estimation of $C(x)$ it is convenient to work with its periodic form
\begin{eqnarray}
C(x)&=&\int_{-\pi/2}^{\pi/2}p(x-y)\Theta(y) dy,\label{eq:cdf}
\end{eqnarray}
where $\Theta(x)$ is a $2\pi$ periodic Heaviside function
\begin{equation*}
\Theta(x)=\begin{cases}
            1, & \quad x\in [2k\pi,(2k+1)\pi) \\
            0 & \quad x\in  [(2k-1)\pi,2k\pi),
          \end{cases}
\end{equation*}
where $k\in \mathbb{Z}$. It is straightforward to check that this definition results in the desired $C(x)$ as given in Eqn.~(\ref{eq:cdf1}).

One then proceeds by considering a Fourier-series approximation of $\Theta(x)$
\begin{eqnarray}
F(x)&=&\sum_{k\in S_1} F_k e^{ikx},\label{eq:fourier}
\end{eqnarray}
where $S_1=\{0\}\bigcup \{\pm (2j+1)\}_{j=0}^d$, $d\in \mathbb{N}$ and the Fourier coefficients $F_j$ are
\begin{align}
& F_k=\begin{cases}
& \frac{1}{2}, \quad k=0 \\
&-i\sqrt{\frac{\beta}{2\pi}}e^{-\beta}\frac{I_j(\beta)+I_{j+1}(\beta)}{2j+1}, \quad k=2j+1 \\ 
&-i\sqrt{\frac{\beta}{2\pi}}e^{-\beta}\frac{I_j(\beta)}{2d+1},\quad k=2d+1,
\end{cases}\label{eq:fourier1}
\end{align}
where  $0\leq j\leq d-1$, $I_n$ is the modified Bessel function of first kind, and 
\begin{equation}
F_{-(2j+1)}=-F_{2j+1}, \quad 0< j\leq d.\label{eq:fourier4}
\end{equation}
%The authors carefully constructed this approximation by truncating the Chebyshev polynomial expansion for an $erf(·)$ function to satisfy 
As shown in \cite{spe}, the above Fourier approximation satisfies 
\begin{equation*}
|\Theta(x)-F(x)|\leq \epsilon, \forall x\in [-\pi+\delta,-\delta]\cup[\delta,\pi-\delta],
\end{equation*}
provided $d$ and $\beta$ are selected as 
\begin{equation}
d=O(\frac{1}{\delta}\log\frac{1}{\epsilon}), \quad \beta=\max\left\{\frac{1}{4 \sin^2\delta}W\left(\frac{3}{\pi \epsilon^2}\right),1\right\},\label{eq:forpara}
\end{equation}
where $W(\cdot)$ denotes the principal branch of the Lambert-W function and $\delta\in (0,\pi/2)$. Furthermore
\begin{align*}
&\sum_{k\in S_1}|F_k|=O(\log d), \quad -\epsilon \leq |F(x)|\leq 1+\epsilon, \forall x\in \mathbb{R}.
\end{align*}   
Given this Fourier approximation of $\Theta(x)$, approximate CDF (ACDF) is defined as
\begin{align*}
&\tilde{C}(x)= \int_{-\pi/2}^{\pi/2} p(y) F(x-y)dy =\sum_{j\in S_1}F_j\int_{-\pi/2}^{\pi/2} p(y)e^{ij(x-y)}dy \\
&=\sum_{j\in S_1}F_je^{ijx}\text{Tr}[\rho e^{-ij\tau H}], 
\end{align*}
which satisfies
\begin{equation}
C(x-\delta)-\epsilon\leq \tilde{C}(x)\leq C(x+\delta)+\epsilon.\label{eq:cineq}
\end{equation}
 
\subsection{Generalized Randomized Compilation Lemma}
In this section we generalize the random compilation lemma (Lemma 2) from \cite{spe} to handle arbitrary Pauli weights $\alpha_i$ in the LCU decomposition of $H$. Define 
\begin{eqnarray}
% \nonumber to remove numbering (before each equation)
\hat{H} &=& \frac{1}{\lambda}\sum_{i=1}^{L}|\alpha_i|\text{sgn}(\alpha_i)P_i=\sum_{i=1}^{L}p_i\tilde{P}_i,\label{eq:hhat}
\end{eqnarray}
where $\tilde{P}=\text{sgn}(\alpha_i)P_i$ and $p_i=\frac{|\alpha_i|}{\lambda}>0$. Note that $\sum_{i=1}^{L}p_i=1$ and
\begin{equation*}
H=\lambda \hat{H}.
\end{equation*}
Then
\begin{align*}
% \nonumber to remove numbering (before each equation)
  &e^{i\hat{H}x} = \sum_{n} \frac{(i\hat{H}x)^n}{n!}=\sum_{n \text{ even}} \frac{(i\hat{H}x)^n}{n!}+\frac{(i\hat{H}x)^{n+1}}{(n+1)n!}\\
  &= \sum_{n \text{ even}} \frac{(i\hat{H}x)^n}{n!}\left(I+\frac{i\hat{H}x}{(n+1)}\right) \\
  &= \sum_{n \text{ even}} \frac{(ix)^n}{n!}\left(\sum_{i=1}^{L}p_i\tilde{P}_i\right)^n\left(I+\frac{ix\sum_{j=1}^{L}p_j\tilde{P}_j }{(n+1)}\right) \\
  &= \sum_{n \text{ even}} \frac{(ix)^n}{n!}\left[\left(\sum_{l_1,l_2,\cdots,l_n}p_{l_1}p_{l_2}\cdots p_{l_n}\tilde{P}_{l_1}\tilde{P}_{l_2}\cdots \tilde{P}_{l_n}\right)\right.\\
  &\quad \quad\quad\quad\quad\quad \left.\times\left(I\sum_{j=1}^{L}p_j+\frac{ix\sum_{j=1}^{L}p_j\tilde{P}_j }{(n+1)}\right)\right]\\
  &= \sum_{n \text{ even}} \frac{(ix)^n}{n!}\left[\left(\sum_{l_1,l_2,\cdots,l_n}p_{l_1}p_{l_2}\cdots p_{l_n}\tilde{P}_{l_1}\tilde{P}_{l_2}\cdots \tilde{P}_{l_n}\right)\right.\\
  & \quad \quad\quad\quad\quad\quad \left.\times \sum_{l^{\prime}}p_{l^{\prime}}\left(I+\frac{ix}{(n+1)}\tilde{P}_{l^{\prime}}\right)\right]\\
  &= \sum_{n \text{ even}} \frac{(ix)^n}{n!}\sqrt{1+(\frac{x}{n+1})^2}\left[\sum_{l_1,l_2,\cdots,l_n,l^{\prime}}p_{l_1}p_{l_2}\cdots p_{l_n}p_{l^{\prime}}\right.\\
  &\quad \quad\quad\quad\quad\quad \quad \quad\quad\quad\quad\quad\left.\times (\tilde{P}_{l_1}\tilde{P}_{l_2}\cdots \tilde{P}_{l_n}V^n_{l^{\prime}})\right].
\end{align*}
where
\begin{align*}
&V^n_{l^{\prime}}=\frac{1}{\sqrt{1+(\frac{x}{n+1})^2}}\left(I+\frac{ix}{(n+1)}\tilde{P}_{l^{\prime}}\right)\\
&=\cos \theta_n+i\sin \theta_n \tilde{P}_{l^{\prime}}=e^{i\theta_n \tilde{P}_{l^{\prime}}},
\end{align*}
and
\begin{equation*}
\theta_n=\text{sgn}(x)\text{arccos}\left(\frac{1}{\sqrt{1+(\frac{x}{n+1})^2}}\right).
\end{equation*}
Thus, for given $t$ and integer $r\geq 0$
\begin{align}
% \nonumber to remove numbering (before each equation)
&e^{i\hat{H}t}=e^{(i\hat{H}t/r)^r}=(\sum_{m}c_m W_m)^r\notag\\
&=\sum_{m_1 m_2\cdots m_r}c_{m_1}c_{m_2}\cdots c_{m_r} W_{m_1}W_{m_2}\cdots W_{m_r}=\sum_{k\in S_2} b_k U_k, \label{eq:complemma}
\end{align}
for some index set $S_2$ where 
\begin{equation*}
b_k=c_{m_1}c_{m_2}\cdots c_{m_r},\quad U_k=W_{m_1}W_{m_2}\cdots W_{m_r}, 
\end{equation*}
with 
\begin{equation*}
c_{m_j}=\frac{|t/r|^{n_j}}{n_i!}\sqrt{1+\left(\frac{ t/r}{n_j+1}\right)^2} p_{l_1}p_{l_2}\cdots p_{l_{n_j}}p_{l^{\prime}}>0,
\end{equation*}
\begin{align}
&W_{m_j}=(i\text{sgn}(t))^{n_j} \tilde{P}_{l_1}\tilde{P}_{l_2}\cdots \tilde{P}_{l_{n_j}} e^{i\theta_{n_j} \tilde{P}_{l^{\prime}}},\notag\\
&=(i\text{sgn}(t))^{n_i} \text{sgn}(\alpha_{l_{1}}\cdots \alpha_{l_{n_j}}) P_{l_1}P_{l_2}\cdots P_{l_{n_j}} e^{i\theta_{n_j} \text{sgn}(\alpha_{l^{\prime}})P_{l^{\prime}}}\notag,
\end{align}
and
\begin{equation*}
\theta_{n_j}=\text{sgn}(t)\,\text{arccos}\!\left(\frac{1}{\sqrt{1+(\frac{t/r}{n_j+1})^2}}\right), 
\end{equation*}
for $n_j,j=1,\cdots,r$ even integers.  Note that for a given $n_j$, indices $m_j$ above correspond to specific choices of $l_1,\cdots,l_k,\cdots,l_{n_j}$ where $1\leq l_{k}\leq L, k=1,\cdots,n_j$. Also note
\begin{equation*}
\sum_{k\in S_2}b_k=\left(\sum_{m}c_m\right)^{r},
\end{equation*}
and $b_k>0$ by construction.  Furthermore
\begin{equation}
\sum_{m}c_m=\sum_{n=0}^{\infty}\frac{1}{(2n)!}(t/r)^{2n}\sqrt{1+\left(\frac{t/r}{2n+1}\right)^2},\label{eq:Csum1}
\end{equation}
which can upper bounded as follows
\begin{equation}
\sum_{m}c_m\leq \exp{(t^2/r^2)}, \label{eq:Csum2}
\end{equation}
as shown in \cite{spe}.

To sample a term $U_k$ in (\ref{eq:complemma}) one can use a Monte Carlo sampling procedure described in Algo.~\ref{algo:sampleU}. 
\begin{algorithm}
\caption{Sampling $U$}\label{algo:sampleU}
\label{alg:binary_search}
\begin{algorithmic}[1]
\State \textbf{Inputs}: $t,r,\{\alpha_j P_j\}_{j=1}^L$ and tolerance $\epsilon_q>0$.
\State Initialize $U=I,s=1$
\For{$j=1:r$}
\State Sample an even integer $n$ with probability
\begin{equation*}
q_n \propto \frac{(t/r)^n}{n!}\sqrt{1+\left(\frac{t/r}{n+1}\right)^2},
\end{equation*}
where one can use a truncated distribution as follows. Compute $q^t_j=q_{2j}$ for $j=1,2,\cdots,N$ such that $Q=\sum_{j=1}^{N}q^t_j\leq 1-\epsilon_q$, and then sampling $n=2j$ from that finite distribution $i\sim \{q^t_j/Q\}_{j=1}^n$. 
\State Independently sample $n+1$ indices $l_1,l_2,\cdots,l_n, l^{\prime}$ from $\{p_l=\frac{|\alpha_i|}{\lambda}\}_{l=1}^L$
\State Compute 
\begin{equation*}
\theta_{n}=\text{sgn}(t)\,\text{arccos}\!\left(\frac{1}{\sqrt{1+(\frac{t/r}{n+1})^2}}\right).
\end{equation*}
\State Compute $U=P_{l_1}P_{l_2}\cdots P_{l_n} e^{i\theta_n \text{sgn}(\alpha_{l^{\prime}})P_{l^{\prime}}} U$
\State Compute $s=s\times (i \times \text{sgn}(t))^n \text{sgn}(\alpha_{l_1}\alpha_{l_2}\cdots \alpha_{l_n})$ 
\EndFor
\State \textbf{Return}: $s\times U$
\end{algorithmic}
\end{algorithm}

\subsection{Exploiting Symmetry}
In this section we utilize the symmetry of Fourier coefficients (\ref{eq:fourier1}-\ref{eq:fourier4}) to simply computation of the ACDF $\tilde{C}(x)$. Let $t_j=-j\tau\lambda$ then 
\begin{align*}
&\tilde{C}(x)= \sum_{j\in S_1}F_je^{ijx}\text{Tr}[\rho e^{-ij\tau H}]=\sum_{j\in S_1}F_je^{ijx}\text{Tr}[\rho e^{it_j \hat{H}}],  
\end{align*}
where $\hat{H}$ is as defined in (\ref{eq:hhat}). Lets decompose $S_1=\{0\}\bigcup S_1^{+}\bigcup S_1^{-}$ where $S_1^{\pm}=\{\pm(2j+1);j=1,\cdots,d\}$. From definition of $F_j$ in (\ref{eq:fourier1}) and the property (\ref{eq:fourier4}), it is clear that $F_j=-i|F_j|, j>0$ and $F_j=i|F_j|, j<0$. Also note that $z_{j}=\text{Tr}[\rho e^{it_j \hat{H}}]=\overline{z}_{-j}$. Thus 
\begin{align*}
&\tilde{C}(x)=\frac{1}{2}-i\sum_{j\in S_1^+}|F_j|e^{ijx}\text{Tr}[\rho e^{it_j \hat{H}}]+i\sum_{j\in S_1^-}|F_j|e^{ijx}\text{Tr}[\rho e^{it_j \hat{H}}],\\
&=\frac{1}{2}-i\sum_{j\in S_1^+}|F_j|e^{ijx}\text{Tr}[\rho e^{it_j \hat{H}}]+i\sum_{j\in S_1^+}|F_j|e^{-ijx}\text{Tr}[\rho e^{-it_j \hat{H}}]\\
&=\frac{1}{2}+\sum_{j\in S_1^+}i|F_j|\left(\overline{e^{ijx}\text{Tr}[\rho e^{it_j \hat{H}}]}-e^{ijx}\text{Tr}[\rho e^{it_j \hat{H}}]\right)\\
&=\frac{1}{2}+2\sum_{j\in S_1^+}|F_j|\text{Im}\left(e^{ijx}\text{Tr}[\rho e^{it_j \hat{H}}]\right)\\
&=\frac{1}{2}+2\tilde{C}_p(x)
\end{align*}
where
\begin{align}
&\tilde{C}_p(x)=\sum_{j\in S_1^+}|F_j|\left(\sin(jx)\text{Re}(\text{Tr}[\rho e^{it_j \hat{H}}])\right.\notag\\
&\qquad\qquad\qquad+\left.\cos(jx)\text{Im}(\text{Tr}[\rho e^{it_j \hat{H}}])\right).
\end{align}
Finally, using Eqn. (\ref{eq:complemma}) with $t=t_j$ we get
\begin{align*}
% \nonumber to remove numbering (before each equation)
&\tilde{C}_p(x)=\sum_{j\in S_1^+}\sum_{k\in S_2}|F_j|b^j_k\Gamma_{j,k}(x),
\end{align*}
where
\begin{equation}
\Gamma_{j,k}(x)=\left(\sin(jx)\text{Re}(\text{Tr}[\rho U^j_k])+\cos(jx)\text{Im}(\text{Tr}[\rho U^j_k])\right).\label{eq:gamma}
\end{equation}
Define 
\begin{equation*}
a_{jk}=b^j_k|F_j|,
\end{equation*}
where recall $b^j_k>0$. Select a runtime vector $\overrightarrow{r}=(r_1,\cdots,r_{|S_1^+|})$ of natural numbers (specific choice given in next section) and let 
\begin{align}
A(\overrightarrow{r})&=\sum_{j\in S_1^{+}, k\in S_2} |a_{jk}|=\sum_{j\in S_1^+,} |F_j|\left(\sum_{k\in S_2}b^j_k\right)=\sum_{j\in S_1^+} |F_j|\mu_j,  \label{eq:Adef}
\end{align}
where
\begin{equation*}
\mu_j=\sum_{k\in S_2}b^j_k=(\sum_{m}c_m)^{r_j}, 
\end{equation*}
using the relation (\ref{eq:Csum1}). Finally, note that 
\begin{eqnarray*}
\tilde{C}(x) &=&\frac{1}{2}+2A(\overrightarrow{r}) \sum_{j\in S_1^{+}, k\in S_2}\frac{a_{jk}}{A(\overrightarrow{r})}\Gamma_{j,k}(x), 
\end{eqnarray*}
and using $P(J=j,K=k)=\frac{a_{jk}}{A(\overrightarrow{r})}$ as a probability distribution one can approximate $\tilde{C}(x)$ via Monte Carlo sampling by noting that
\begin{eqnarray*}
\tilde{C}(x) &=&\frac{1}{2}+2A(\overrightarrow{r}) E_{P}[\Gamma_{j,k}(x)],
\end{eqnarray*}
where $E_P$ is the expectation operator. A sampling procedure to estimate $\tilde{C}(x)$ and an approximate method 
to compute $A(\overrightarrow{r})$ are summarized in the Algo.~\ref{algo:ACDF} and Algo. \ref{algo:A}, respectively. 

Let $\hat{\tilde{C}}_N(x)$ be an empirical estimate of $\tilde{C}(x)$ obtained via $N$ samples, see Appendix~\ref{sec:errprob}.  
In order to detect first jump in the $\tilde{C}(x)$ we assume that
\begin{align}
& \hat{\tilde{C}}_N(x)\geq \eta/2 \Rightarrow \tilde{C}(x)>\epsilon, \label{eq:cond1}\\
& \hat{\tilde{C}}_N(x)< \eta/2 \Rightarrow \tilde{C}(x)<\eta-\epsilon, \label{eq:cond2}
\end{align}
where $\eta\leq p_0$ with $p_0$ being square of overlap with ground state of $H$ and $\epsilon\in (0,\eta/2)$. By using (\ref{eq:cineq}), condition (\ref{eq:cond1}) implies $C(x+\delta)>0$ and the condition (\ref{eq:cond2}) implies $C(x-\delta)<\eta$ . 

There may be error in the assumptions (\ref{eq:cond1}-\ref{eq:cond2}) due to finite sampling. As shown in the Appendix~\ref{sec:errprob}, one can keep the probability of this error bounded, i.e., $p_{err}\leq \nu$ (see Eqn. (\ref{eq:errprob})) for any given $\nu\in [0,1]$ by taking
\begin{equation}
N=N_s(\overrightarrow{r})=\left\lceil 8\left(\frac{A(\overrightarrow{r})}{\eta/2-\epsilon}\right)^2\ln \frac{1}{\nu}\right\rceil,\label{eq:samples}
\end{equation}
samples. We next compare $N_s(\overrightarrow{r})$ with \#samples $C^o_s(\overrightarrow{r})$ derived in \cite{spe} given by 
\begin{equation}
N_s^o(\overrightarrow{r})=\left\lceil \left(\frac{2A^o(\overrightarrow{r})}{\eta/2-\epsilon}\right)^2\ln \frac{1}{\nu}\right\rceil,\label{eq:samplesorg}
\end{equation}
where 
\begin{align*}
A^o(\overrightarrow{r})&=\sum_{j\in S_1} |F_j|\mu_j.
\end{align*}
Note that sum in $A^o(\overrightarrow{r})$ runs over $S_1$ instead of $S^+_1$ in $A(\overrightarrow{r})$. Assuming $\overrightarrow{r}$ to be as given in (\ref{eq:rchoice}), it is straightforward to show that 
\begin{align}
A^o(\overrightarrow{r})&=2A(\overrightarrow{r})+\frac{1}{2}.\label{eq:Aodef}
\end{align}
Substituting (\ref{eq:Aodef})  in (\ref{eq:samplesorg}) it follows that $N_s(\overrightarrow{r})<\frac{1}{2}N_s^o(\overrightarrow{r})$.   Thus by exploiting symmetry of Fourier coefficients in computing ACDF, one can reduce \#samples by factor of $1/2$ while keeping the same error probability and accuracy in computation of ACDF. While above analysis assumed particular choice of $\overrightarrow{r}$, this result holds in general. 

\section{Modified SPE Algorithm}\label{sec:spealgo}
Given the derivations in previous section, we next summarize modified QSPE algorithm.
The inputs are: 
\begin{itemize}
  \item LCU decomposition of the Hamiltonian $H=\sum_{j=1}^{L}\alpha_j P_j$.
  \item Trial state $\rho$.
  \item $\Delta>0$, the precision in estimate so that $|\beta_0-\tilde{\beta}_0|\leq \Delta$. 
  \item Lower bound $\eta\in (0,1]$ on overlap of $\rho$ with ground-state energy, so that $p_0=\text{Tr}[\rho\Pi_0]\geq \eta$. 
  \item Probability of error $\nu\in (0,1)$ in computation of GSE.
\end{itemize}
Additionally, following parameters need to be selected:
\begin{itemize}
  \item Select $\epsilon\in (0,\eta/2)$.
  \item Select $\delta\in (0,\min(\pi/2,\tau\Delta)]$, where $\lambda=\sum_{j=1}^{L}|\alpha_j|$ and $\tau=\frac{\pi}{2\lambda+\Delta}$.
  \item Select $d,\beta$ as per (\ref{eq:forpara}).
  \item Select $\epsilon_q,\epsilon_c \ll 1$ which determines approximation accuracies in Algo.~\ref{algo:sampleU} and Algo.~\ref{algo:A}, respectively.
  \item Select the runtime vector $\overrightarrow{r}=(r_1,\cdots,r_{|S^+_1|})$. Once choice is 
  \begin{align}
  r_j=\lceil 2 t^2_j\rceil.\label{eq:rchoice}
  \end{align}
  An optimization procedure to select $\overrightarrow{r}$ is discussed in Section~\ref{sec:ropt}. 
\end{itemize}

Algorithm~\ref{alg:binary_search} describes a binary search to estimate GSE $\tilde{\beta}_0$ such that $|\beta_0-\tilde{\beta}_0|\leq \Delta$. 
As shown in \cite{lin2022heisenberg} it takes 
\begin{align}\label{eq:iter}
N_{iter}=O\left(\log \frac{1}{\delta}\right),
\end{align}
iterations for this search to converge.  At each iteration, $\tilde{C}(x)$ is estimated using Algo.~\ref{algo:ACDF} which involves two Hadamard tests in the Step~\ref{step:stepHar} and is the only quantum step.  Note that one can reuse the samples collected in this step for all of the different $x$ values, with only a small overhead in the sample complexity.  The Hadamard test involves a sequence of controlled multi-qubit Pauli operator and controlled multi-qubit Pauli rotations based on sampling in Algo.~\ref{algo:sampleU}. Given a multi-qubit Pauli operator $P$, controlled Pauli rotation $e^{i\theta P}$ can be implemented via noting
\begin{equation*}
\ket{0}\bra{0}\otimes I+\ket{1}\bra{1}\otimes e^{i\theta P}=e^{i\frac{\theta}{2}I\otimes P}e^{-i\frac{\theta}{2}Z_0\otimes P},
\end{equation*}
where $Z_0$ is Pauli-Z gate applied to the control qubit, and using standard circuit for implementing $e^{i\theta P}$ in terms of $CX$ (CNOT) , $H$ (Hadamard), $S$ (Phase) and $e^{i\theta X}$ (single qubit Pauli-X rotation) gates \cite{sarkar2024scalable}. Further simplifications in compiling circuit for Hadamard test are possible as discussed in \cite{spe}.  

\begin{algorithm}
\caption{Binary Search for GSE Estimation}\label{alg:binary_search}
\begin{algorithmic}[1]
\State \textbf{Inputs}: $\eta, \tau, \delta$
\State Set $l=0$, $x_{0,0}=-\pi/2$ and $x_{1,0}=\pi/2$.
\While{$x_{1,l}-x_{0,l}> 2 \delta$ }
\State $x_l=\frac{1}{2}(x_{1,l}+x_{0,l})$
\State Compute $c=\tilde{C}(x_l)$ using Algo.~\ref{algo:ACDF}
\If{$c<\eta/2$}
\State Set flag=0
\Else
\State Set flag=1
\EndIf
\If{flag}
\State $x_{1,l+1}=x_l+\frac{2}{3}\delta$ and $x_{0,l+1}=x_{0,l}$.
\Else
\State $x_{1,l+1}=x_{1,l}$ and $x_{0,l+1}=x_{l}-\frac{2}{3}\delta$.
\EndIf
\State $l\leftarrow l+1$
\EndWhile
\State $x^*=\frac{1}{2}(x_{1,l}+x_{0,l})$
\State \textbf{Return}: $x^*/\tau$
\end{algorithmic}
\end{algorithm}

\begin{algorithm}
\caption{Computation of $\tilde{C}(x)$}\label{algo:ACDF}
\begin{algorithmic}[1]
\State \textbf{Inputs}: $x,\overrightarrow{r},\rho,\eta,\epsilon,\nu,\Delta, \{\alpha_j,P_j\}_{j=1}^L, S_1^+, \{F_j\}_{j\in S_1^+}$
\State Compute $\mathcal{F}=\sum_{j\in S_1^+}|F_j|$, $\lambda=\sum_{j=1}^{L}|\alpha_j|$ and $\tau=\frac{\pi}{2\lambda+\Delta}$
\State Compute $A(\overrightarrow{r})$ using Algo.~\ref{algo:A}.
\State Compute \#samples $N_s(\overrightarrow{r})$ using Eqn.~(\ref{eq:samples})
\For{$n=1:N_s$}
\State Sample $j\sim \frac{|F_j|}{\mathcal{F}}$. 
\State Sample $U^j_k$ using Algo.~\ref{algo:sampleU} with inputs $r_j$ and $t_j=-j\tau\lambda$.
\State Run Hadamard tests to compute $z^j_r=\text{Re}(\text{Tr}[\rho U^j_k])$ and $z^j_{im}=\text{Im}(\text{Tr}[\rho U^j_k])$.\label{step:stepHar}
\State $z_n=A(\overrightarrow{r})\left(\sin(jx)z^j_r+\cos(jx)z^j_{im})\right)$.
\EndFor 
\State Compute $\overline{z}=\frac{1}{2}+2\frac{\sum_{n}z_n}{N_s} $
\State \textbf{Return}: $\overline{z}$
\end{algorithmic}
\end{algorithm}

\begin{algorithm}
\caption{Computation of $A(\overrightarrow{r})$}\label{algo:A}
\begin{algorithmic}[1]
\State \textbf{Inputs}: $\tau,\lambda,S_1^+$, $\{F_j\}_{j\in S_1^+}$, $\overrightarrow{r}=(r_1,\cdots,r_{|S_1^+|})$ and tolerance $\epsilon_c>0$
\State Set $A(\overrightarrow{r})=0$.
\For{$j\in S_1^+$}
\State Compute $t_j=-j\tau\lambda$
\State Compute approximation for $C_j$ as
\begin{align*}
&C_j=\sum_{n=0}^{\infty}\frac{1}{(2n)!}(t_j/r_j)^{2n}\sqrt{1+\left(\frac{t_j/r_j}{2n+1}\right)^2}\\
&\approx \sum_{n=0}^{N}\frac{1}{(2n)!}(t_j/r_j)^{2n}\sqrt{1+\left(\frac{t_j/r_j}{2n+1}\right)^2},
\end{align*}
for some large even integer $N$ such that $\frac{1}{(2N)!}(t_j/r_j)^{2N}\sqrt{1+(\frac{t_j/r_j}{2N+1})^2}<\epsilon_c$.
\State Compute $\mu_j=(C_j)^{r_j}$
\State $A(\overrightarrow{r})=A(\overrightarrow{r})+|F_j|\mu_j$
\EndFor
\State \textbf{Return}: $A(\overrightarrow{r})$
\end{algorithmic}
\end{algorithm}

The expected number of controlled Pauli rotations per Hadamard test circuit is given by
\begin{align}\label{eq:gatecomp}
N_g(\overrightarrow{r}) &= \frac{1}{A(\overrightarrow{r})}\sum_{(j,k)\in S_1^{+}\times S_2}|a_{jk}|r_j.
\end{align}
Since Step~\ref{step:stepHar} in Algo.~\ref{algo:ACDF} is repeated $N_s(\overrightarrow{r})$ (see Eqn. (\ref{eq:samples})) times, so the expected total non-Clifford complexity is $2N_s(\overrightarrow{r})N_g(\overrightarrow{r})$. Given binary search takes $N_{iter}$ (Eqn.~(\ref{eq:iter})) iterations to converge, total non-Clifford complexity of entire algorithm is $2N_s(\overrightarrow{r})N_g(\overrightarrow{r})N_{iter}$. 

If one is exclusively interested in asymptotic complexity of QSPE algorithm, the simple choice for $\overrightarrow{r}$ in Eqn. (\ref{eq:rchoice}) implies $N_g(\overrightarrow{r})\leq \max_{j\in S_1^+} r_j\leq 2 \lceil (2d+1)\lambda \tau \rceil^2$ and $A(\overrightarrow{r})\leq \sqrt{e}\sum_{j\in S_1^+}|F_j|=O(\log d)$, using which leads to
\begin{align*}
N_g(\overrightarrow{r}) &=O\left(\frac{1}{\delta^2}\log^2 \frac{1}{\eta}\right), \\
N_s(\overrightarrow{r}) &=O\left(\frac{1}{\eta^2}\log\frac{1}{\nu}\log^2\!\left(\frac{1}{\delta}\log \frac{1}{\eta}\right)\right),
\end{align*}
where we have taken $\epsilon=O(\eta)$. Since Algo.~\ref{algo:ACDF} errs with probability at most $\nu$ for any $x$, choosing $\nu = \zeta/N_{iter}$ would ensure, by the union bound, that the ground state is successfully estimated with probability at least $1-\zeta$. Thus, taking $\delta=\tau\Delta\sim O(\frac{\Delta}{\lambda})$,  to estimate GSE with additive error $\Delta$ and success probability $1-\zeta$ requires
\begin{itemize}
  \item running  $2N_{iter}N_s(\overrightarrow{r})$  circuits that scales as
  \begin{equation*}
  O\left(\frac{1}{\eta^2}\log^2\left(\frac{\lambda}{\Delta}\log \frac{1}{\eta}\right)\log\left(\frac{1}{\zeta}\log\frac{\lambda}{\Delta} \right)\log\frac{\lambda}{\Delta}\right),
  \end{equation*}
  \item and each circuit has a non-Clifford gate complexity that scales as
  \begin{equation*}
    O\left(\frac{\lambda^2}{\Delta^2}\log^2 \frac{1}{\eta} \right),
  \end{equation*}
\end{itemize}
similar to Theorem 1 in \cite{spe}.

\section{Other Considerations}\label{sec:prac}

\subsection{Optimization of $\overrightarrow{r}$}\label{sec:ropt}
When finding the optimal runtime vector $\overrightarrow{r}$, which determines the sample and gate complexities, one can consider different optimization formulations~\cite{spe}:
\begin{itemize}
  \item Formulation 1: Finding $\overrightarrow{r}$ that minimises the expected total gate complexity, i.e.
  \begin{equation}\label{eq:opt1}
  \min_{\overrightarrow{r}} N_s(\overrightarrow{r})N_g(\overrightarrow{r}).
  \end{equation}
  \item Formulation 2: Finding $\overrightarrow{r}$  that minimises the sample complexity given an upper bound $b_g$ on the expected gate complexity per sample, i.e., 
  \begin{align}
  & \min_{\overrightarrow{r}} N_s(\overrightarrow{r})\label{eq:opt2}\\
  & N_g(\overrightarrow{r})\leq b_g.\label{eq:opt21}
  \end{align}
\end{itemize}
The second formulation is particularly useful in context of EFTQC where it might be advantageous to run a larger number of shorter circuits, even if this increases the total complexity. Furthermore, even though $\overrightarrow{r}\in N^{|S_1^+|}$  is a high dimensional vector since $|S_1^+|\sim d=O(\frac{1}{\delta}\log\frac{1}{\epsilon})$,  as shown in Appendix D \cite{spe} the optimization problems above can be reduced to an efficiently solvable one-dimensional problems.

\subsection{Unknown $\eta$}\label{sec:etabound}
As discussed above, SQPE binary search requires knowledge of $\eta$ that describes the size of the first jump corresponding to GSE. In practice, it is difficult to obtain an estimate of $p_0$ and thereby appropriately select a tight bound $\eta\leq p_0$. On the other hand,  setting $\eta$ to be too small could result in very large number of samples $N_s$ as per the relation~(\ref{eq:samples}). Furthermore, since the number of jumps grows with the size of Hamiltonian (as driven the size of the support of the trial state in the eigenbasis) and the CDF becomes quasicontinuous, jump detection can become difficult to apply. In this section we explore changepoint detection (CD) techniques for identifying GSE which does not rely on knowledge of $\eta$ and can handle quasicontinuous CDF.

Given a sequence of data, a changepoint is a sample or time instant at which some statistical property of a signal changes abruptly. The property in question can be the mean of the signal, its variance, or a spectral characteristic, among others. A variety of methods have been developed for single and multiple CD, see \cite{truong2020selective,aminikhanghahi2017survey} for detailed reviews. 

Let $Y=\{y_k=\tilde{C}(x_k):k=1,\cdots,M\}$ be ACDF samples computed at preselected grid points $X=\{x_k\in [-\pi/2,\pi/2]: k=1,\cdots,M\}$. For instance, given $\Delta$ one could pick $x_k=-\frac{\pi}{2}+(k-1)\Delta, k=1,\cdots,M=\lfloor \frac{\pi}{\Delta}\rfloor+1$. For our application, we are interested in identifying the first changepoint within $Y$ where there is statistically significant abrupt change in the mean of ACDF samples corresponding to the GSE jump. We use binary segmentation which is a search method commonly used within the CD literature. Given $m,n$ with $n\geq m$ we will denote a subsequence of $Y$ by $Y_{m:n}=\{y_k:k=m,m+1,\cdots,n\}$. Define a total mean-square deviation for a subsequence $Y_{m:n}$ as
\begin{equation*}
V(Y_{m:n})=(n-m+1)\text{Var}(Y_{m:n}),
\end{equation*}
where $\text{Var}(Y_{m:n})$ is the variance of data points in $Y_{m:n}$. As outlined in Algo. \ref{algo:CD}, the binary segmentation method uses a single CD method (referred to as $\text{CDAlgo}$) iteratively to detect single changepoint by repeating the method on different subsequences of the sequence $Y$. It begins by initially applying the single CD method to the entire data set (line~\ref{algo:CD1}) and checking the condition defined in line~\ref{algo:CDcond} where $\Delta_V$ is as defined in line~\ref{algo:CDV} and measures difference in variability of the full data subsequence compared to variability in two segments before and after the changepoint $N_c^+$. If $\Delta_V<\Delta_c$ where $\Delta_c$ is small value capturing inherent variability in the sequence, then the changepoint is not significant and the method stops. On the other hand if condition is true, the data is split into two subsequences consisting of the sequence before and after the identified changepoint $M_c^+$, and $\text{CDAlgo}$ is applied to subsequence $Y_{1:M^+_c}$ in line~\ref{algo:CDM}. Finally, the scaled mean of $x$-grid values at the last detected changepoint $x_{M_c^+}/\tau$ and the neighbouring grid point $x_{M_c^+-1}/\tau$ is returned as an estimate of GSE. Note that one can use other approaches, e.g., look for large changes in ACDF derivatives near $x_{M_c^+}$ to estimate GSE as well. For $\text{CDAlgo}$ one can use any method which detects changes in the signal mean. We use the method proposed in \cite{killick2012optimal,lavielle2005using} whose implementation is available in MATLAB.

Finally, note that the CD based approach as presented above requires a priori dense sampling of full ACDF at $M$ grid points unlike the binary search protocol (Algo.~\ref{alg:binary_search}) which adaptively selects such grid points iteratively. Since we are only interested in detecting the first changepoint, one can instead use online CD methods where grid points $x_k$ starting at $-\pi/2$ and corresponding ACDF values $y_k$ are incrementally added to sets $X$ and $Y$ respectively, in a ``streaming fashion" till a changepoint is detected. We would explore such approaches and other stopping criteria based on hypothesis testing in the future work. 

\begin{algorithm}
\caption{Computation of GSE via Changepoint Detection}\label{algo:CD}
\begin{algorithmic}[1]
\State \textbf{Inputs}: $Y=\{y_k=\tilde{C}(x_k):k=1,\cdots,M\}$, $X,\tau$ and $\Delta_c$
\State  Compute $M^+_c=\text{CDAlgo}(Y)$ \label{algo:CD1}
\State  Compute $\Delta_{M_c}=V(Y_{1:M})-(V(Y_{1:M^+_c})+V(Y_{M^+_c+1:M}))$ \label{algo:CDV}
\While{$\Delta_{M_c}>\Delta_c$} \label{algo:CDcond}
\State Compute $M^-_c=\text{CDAlgo}(Y_{1:M^+_c})$\label{algo:CDM}
\State Compute $\Delta_{M_c}=V(Y_{1:M^+_c})-(V(Y_{1:M^-_c})+V(Y_{M^-_c+1:M_c^+}))$
\State $M^+_c\leftarrow M_c^-$
\EndWhile
\State \textbf{Return}: $\frac{x_{M^+_c}+x_{M^+_c-1}}{2\tau}$
\end{algorithmic}
\end{algorithm}

\section{Simulation Results}\label{sec:results}
In this section we apply SQPE for GSE estimation for two model Hamiltonian problems: toy Hamiltonian (Case 1) and molecular dihydrogen (Case 2). The SQPE approach is implemented in the Qiskit environment and simulated using AerSimulator() simulator backend provided by qiskit-aer. For each of these cases we consider trial states with different degree of overlap $\eta$ with the true ground state. For our implementation, we use the choice of $\overrightarrow{r}$ given in Eqn.~(\ref{eq:rchoice}). To assess where this choice falls relative to the optimal $N_s-N_g$ trade-off curve we also apply the second optimization formulation as outlined in the Section~\ref{sec:ropt} for the two cases considered.

We assess the convergence of SQPE approach and report statistics of circuit depth and \# of gates sampled during the binary search (Algo.~\ref{alg:binary_search}) as shown in the Table~\ref{tab:resources}. To get these estimates, the circuits were transpiled to a gate set 
$\{id, rz, sx, x, cx\}$ using Qiskit's built in functions. We also report number of samples used $N_s$ and the absolute error in GSE estimation which we denote by $\Delta_0=\left| \beta_0-\tilde{\beta}_0\right|$. 

\subsection{Case 1: Toy Hamiltonian}
We first consider a simple $3$-qubit Hamiltonian
\begin{equation*}
H=0.2IIZ+0.1ZIX+0.15IZI+0.25IZZ.
\end{equation*}
To generate trial state $\rho=\ket{\phi}\bra{\phi}$ with overlap $\eta=p_0=|\braket{\psi_0|\phi}|^2$ we used
\begin{equation*}
\ket{\phi}=\sqrt{\eta}\ket{\psi_0}+\sqrt{1-\eta}\ket{\psi^{\perp}_0}
\end{equation*}
where $\ket{\psi_0}$ is true ground state and $\ket{\psi^{\perp}_0}$ is a state orthogonal to $\ket{\psi_0}$, i.e., $\braket{\psi^{\perp}_0|\psi_0}=0$. 

\begin{figure}[h!]
  \centering
  \includegraphics[width=\columnwidth]{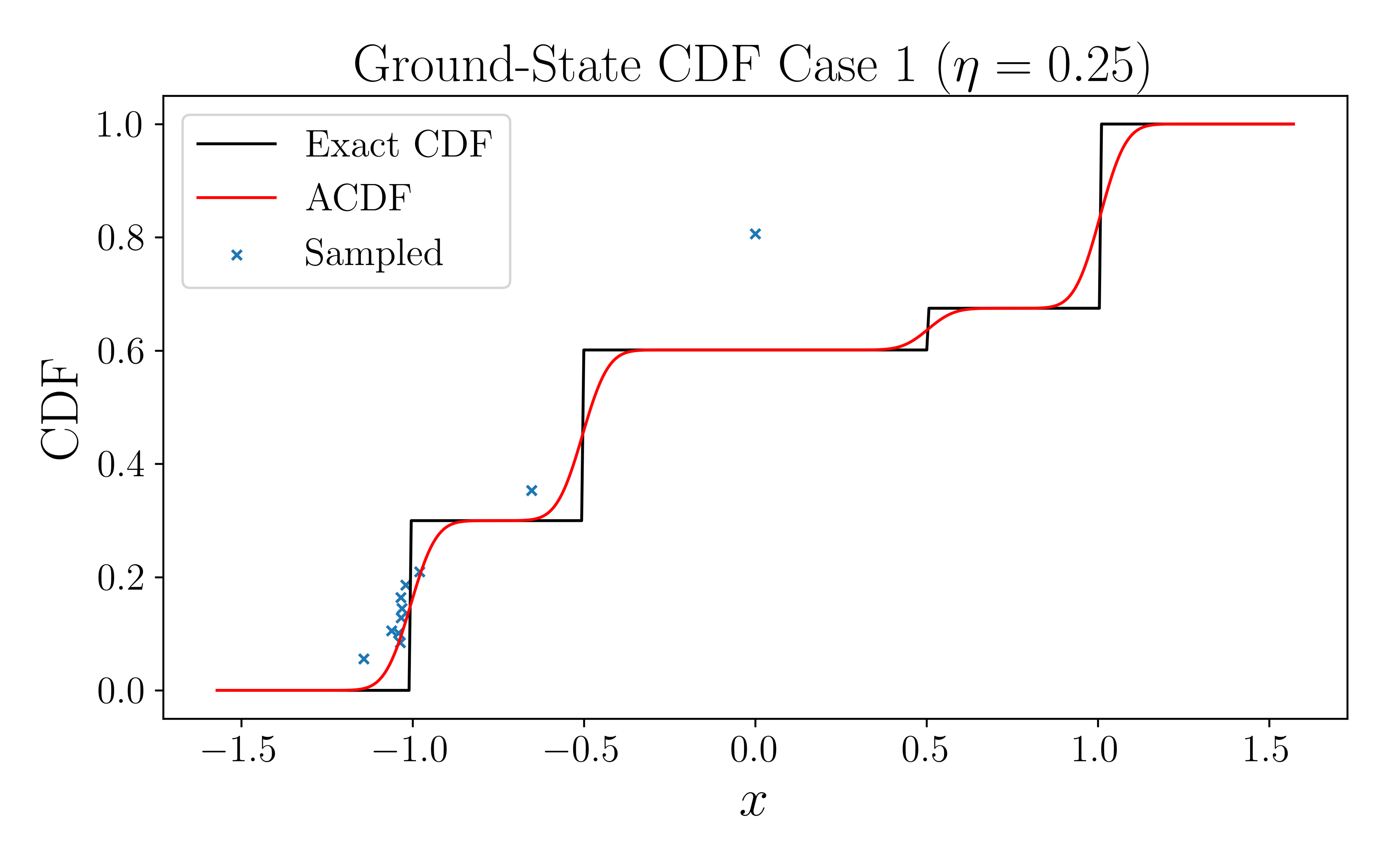}
  \caption{SQPE binary search results for Case 1 with $\eta = 0.25$. Points sampled during the SQPE protocol are shown with the `x' marker, while the black and red curve are the exact CDF and ACDF, respectively. Other algorithm parameters were selected as follows: $\Delta = 0.05$, $\varepsilon=0.1$ and $\nu=0.1$. It takes $N_{iter} = 10$ iterations to converge and the final GSE estimate has an absolute error of $\Delta_0= 0.013$ Ha within prescribed $\Delta$.}\label{fig:case1eta25}
\end{figure}

The resulting qubit Hamiltonian and trial state were used as input to the SQPE protocol. The result of the SQPE binary search (Algo.~\ref{alg:binary_search}) for $\eta=0.25$ is shown in Fig.~\ref{fig:case1eta25}. We can see that the binary search protocol iteratively approaches the true GSE (the x-values cluster around the first jump in the CDF) within the prescribed tolerance $\Delta$. Table~\ref{tab:resources} lists the resource requirements for this case. For comparison, we also include another case with $\eta=0.75$ (with other parameters set as $\Delta = 0.05$, $\varepsilon=\eta/4$ and $\nu=0.1$).
Fig.~\ref{fig:tradeoff} shows the $\tilde{N}_s-N_g$ trade-off curve for $\eta=0.25$, where $\tilde{N}_s=N_s/\log(1/\nu)$ and we have factored out $\log(1/\nu)$ from expression of $N_s$, see Eqn.~(\ref{eq:samples}). Also, shown are $\tilde{N}_s,N_g$ values based on $\overrightarrow{r}$ computed via Eqn.~(\ref{eq:rchoice}), indicating that choice leads to higher $\tilde{N}_s$ value in lieu of smaller circuit, i.e., smaller $N_g$. 
Finally, we apply the CD Algo.~\ref{algo:CD} for $\eta=p_0=0.1$ and $\Delta_c=0.01$. The grid points $X$ were selected uniformly in the range  $[-\pi/2,\pi/2]$ at a resolution of $\Delta=0.057$. The CD Algo. converges in three steps with a GSE absolute error of $\Delta_0=0.034$ Ha. Thus, CD Algo. is able to reliably estimate GSE without requiring the knowledge of $\eta$. 
\begin{figure}[h!]
  \centering
  \includegraphics[width=\columnwidth]{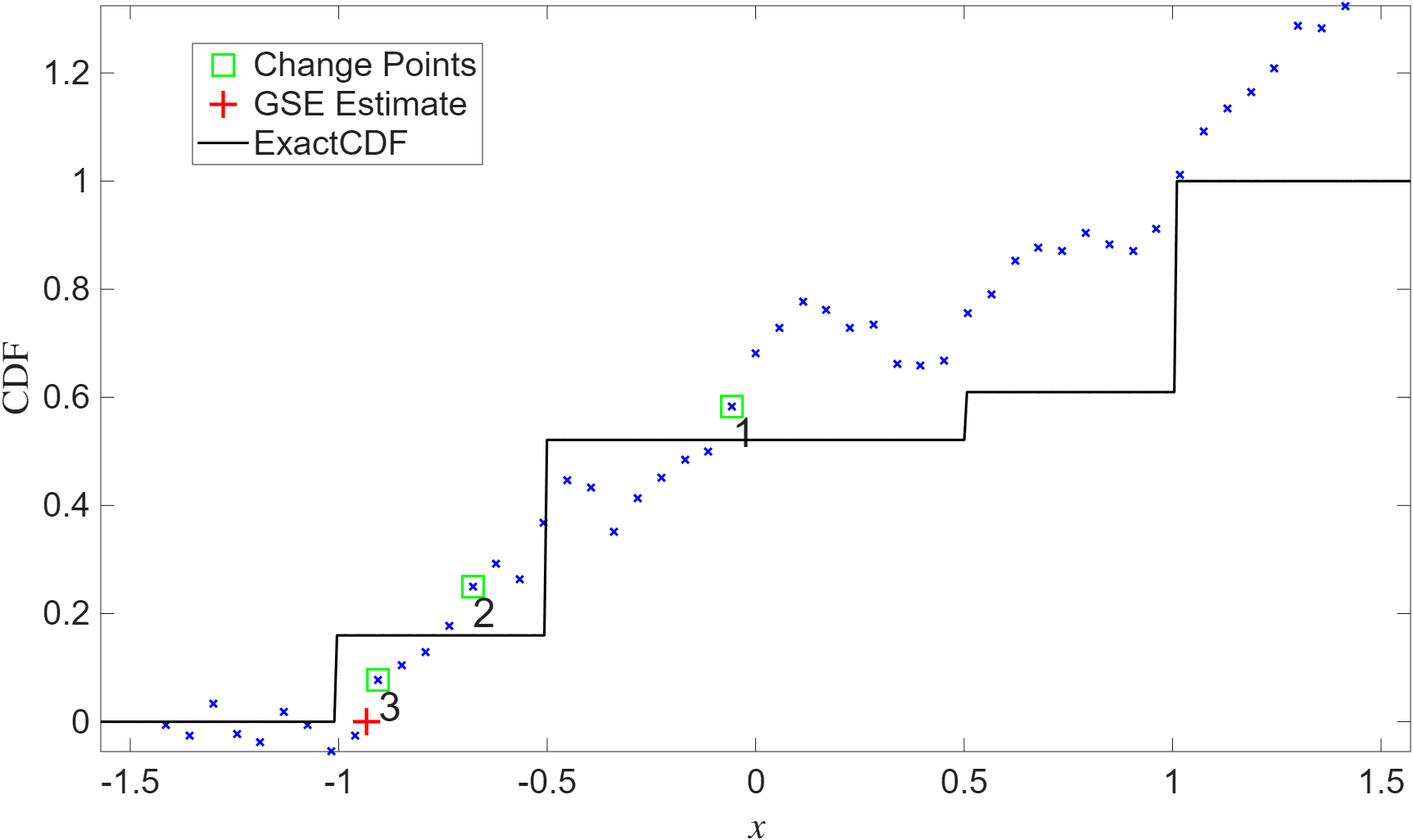} %H2_RSQPE_Search.png
  \caption{CD Algo.~\ref{algo:CD} applied to Case 1 with $\eta=p_0=0.1$. ACDF samples are shown by blue `x' markers, while successive changepoints during the binary segmentation  are shown by green `$\square$' markers. The GSE estimate is shown by red `+' with an absolute error of $\Delta_0=0.034$ Ha. Also shown for reference is true CDF as the black curve.}\label{fig:molecule_fig}
\end{figure}

\subsection{Case 2: H$_2$ Molecule}
To evaluate SQPE on a more realistic model, we consider molecular dihydrogen (H$_2$). The Hamiltonian for H$_2$ was prepared using \texttt{pyscf} and \texttt{OpenFermion} \cite{pyscf, openfermion}. The Hamiltonian was computed for H$_2$ with an interatomic distance of 0.74\ \AA, with the STO-3G basis at the Hartree Fock level of theory requiring $4$-qubits. Once the Fermionic Hamiltonian was obtained, it was converted to the qubit basis using the Jordan-Wigner transformation \cite{jordanwigner} resulting in a LCU with $L=15$ terms and total Pauli weight $\lambda=1.9842$. Some of the Pauli coefficients are negative requiring the extension of random compilation lemma developed in this paper. The trial states with different overlap were generated using a procedure similar to as discussed for the Case 1. 

\begin{figure}[h!]
  \centering
  \includegraphics[width=\columnwidth]{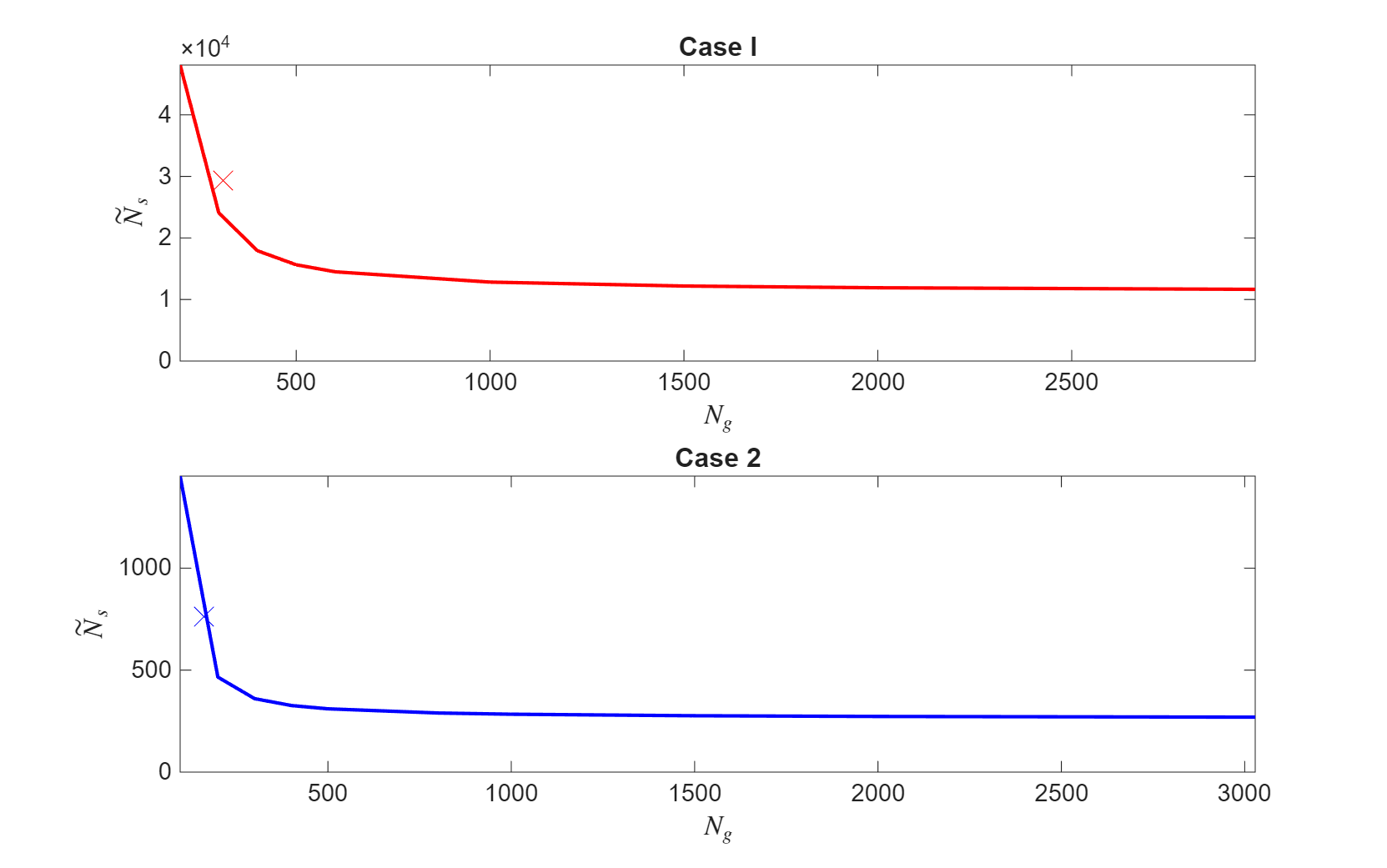}
  \caption{$\tilde{N}_s-N_g$ trade-off curve computed by the optimization formulation Eqn.~(\ref{eq:opt2}-\ref{eq:opt21}) for the Case 1 ($\eta=0.25$) and Case 2 ($\eta=0.5$). The marker `x' shows the $\overrightarrow{r}$ computed via Eqn.~(\ref{eq:rchoice}).}\label{fig:tradeoff}
\end{figure}

The result of the SQPE binary search protocol on H$_2$ for $\eta=0.5$ is shown in Fig.~\ref{fig:molecule_fig}. The estimated GSE iteratively approaches the true GSE within prescribed $\Delta$. Table~\ref{tab:resources} lists the resource requirements for this case and for comparison we also include $\eta=1$ (with all other parameters set to same values as for $\eta=0.5$). Fig.~\ref{fig:tradeoff} shows the $\tilde{N}_s-N_g$ trade-off curve for $\eta=0.5$ with a similar conclusion on the choice of $\overrightarrow{r}$ computed via Eqn.~(\ref{eq:rchoice}) as for the Case 1.   

\begin{figure}[h!]
  \centering
  \includegraphics[width=\columnwidth]{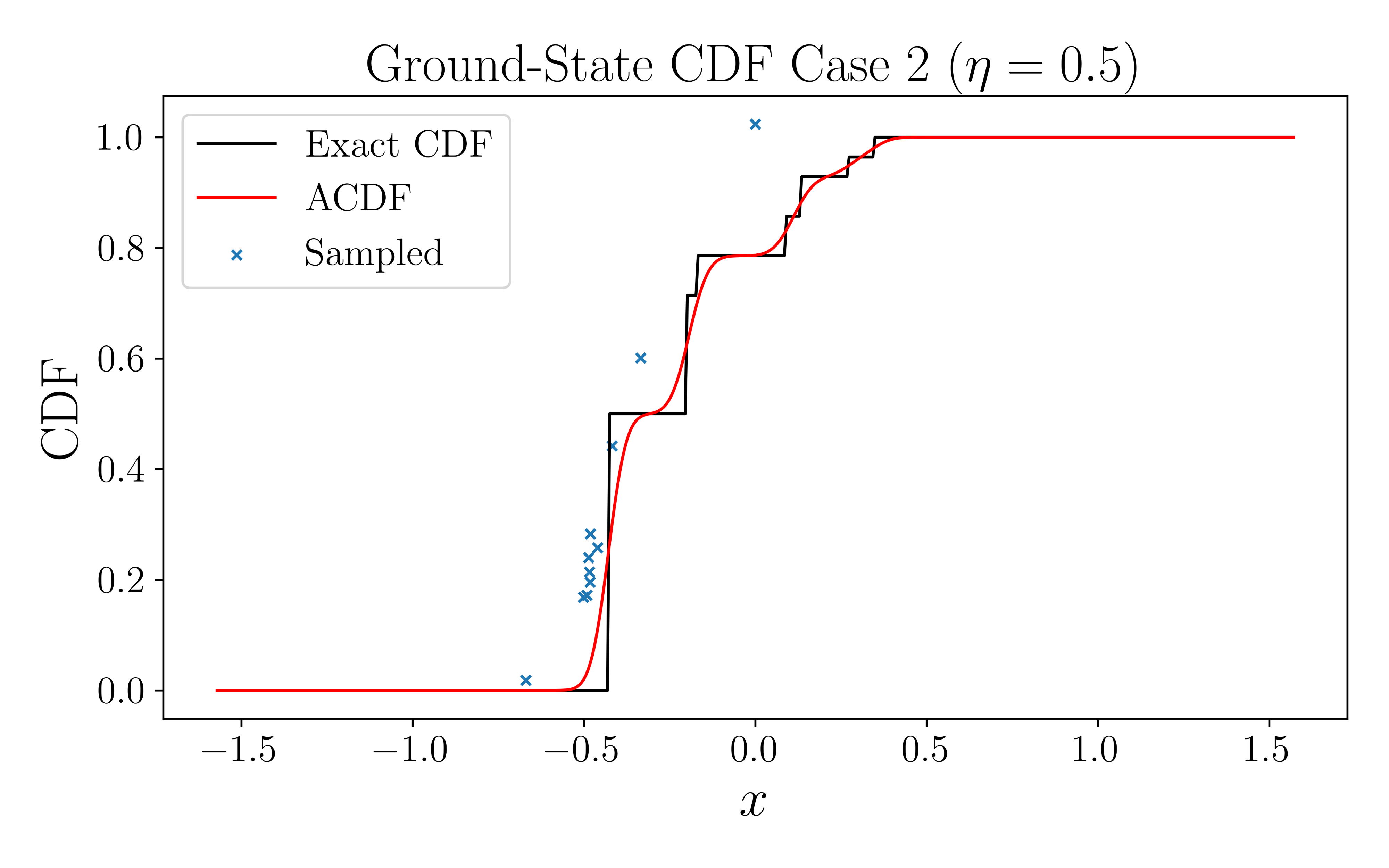} %H2_RSQPE_Search.png
  \caption{Similar to the Fig.~\ref{fig:case1eta25} but for H$_2$ with $\eta = 0.5$. Other algorithm parameters were selected as follows: $\Delta = 0.2$, $\varepsilon=0.1$, and $\nu=0.1$. It took $N_{iter} = 10$ iterations to converge and the final GSE estimate has an absolute error of $\Delta_0= 0.141$ Ha within prescribed $\Delta$.}\label{fig:molecule_fig}
\end{figure}

\begin{table*}
\centering
  \begin{tabular}{|l|l|l|l|l|l|l|l|l|}
    \hline
    \multirow{2}{*}{Case} &
    \multirow{2}{*}{$N_s$} &
    %\multicolumn{1}{c}{$N_s$} &
    \multicolumn{3}{c|}{Circuit Depth} &
    \multicolumn{3}{c|}{Gate Counts} &
    \multirow{2}{*}{$\Delta_0$}\\
     &  & Min & Median & Max & Min & Median & Max & \\ \hline
 Case 1, $\eta=0.75$  & 1500 & 162    &  83,005   & 493,946   &  231   &  111,350  & 662,169  & $0.016$ \\ \hline
 Case 1, $\eta=0.25$  & 1500 &  164   &  104,327    & 462,467  & 229   &  139,777  & 620,000 & $0.013$ \\ \hline
 Case 2, $\eta=1$    & 398 &  70  &   19,859     & 84,027  & 72  &  28,572   & 121,494 & $0.086$ \\ \hline
 Case 2, $\eta=0.5$  & 2825 &  73 & 37,796  & 199,153 &  111   & 55,157 & 289,578 & $0.141$ \\ \hline
  \end{tabular}
\caption{Statistics of circuit depth and gate count sampled during the SQPE binary search protocol for the Cases 1 and 2.  }\label{tab:resources}
\end{table*}

\section{Conclusion}\label{sec:conc}
We developed several refinements and extensions of the SQPE framework to address some of the key practical challenges, improving its applicability to realistic cases. This included generalization of random compilation lemma for dealing with negative Pauli weights in the LCU decomposition, exploiting symmetry of the Fourier expansion to reduce number of circuit runs by half, and exploring changepoint detection method to find GSE without relying on the knowledge of an estimate of the overlap of trial state with the ground state. We illustrated these new developments numerically on two examples in a quantum simulator. 

In future we plan to evaluate SQPE on quantum hardware in conjunction with noise mitigation approaches and explore early fault tolerant implementation with availability of partial error correction. It would also be worthwhile to develop an online changepoint detection approach for improving the sample efficiency for estimating GSE and further generalizing the approach for additionally determining higher excited states by detecting multiple changepoints. 

\appendix

\section{Error Probability}\label{sec:errprob}
Let $\Gamma_l$ be $N$  i.i.d. samples of $\Gamma_{j,k}(x)$ (see Eqn.~\ref{eq:gamma}), then empirical estimate of ACDF is given by
\begin{align}
&\hat{\tilde{C}}_{N}(x)=\frac{1}{2}+2\frac{1}{N}(\sum_{l=1}^{N}S_l),
\end{align}
where $S_l=A(\overrightarrow{r})\Gamma_l$. Note that $|S_l|\leq A(\overrightarrow{r})$.
Then based on the conditions (\ref{eq:cond1}-\ref{eq:cond2}) error probability $p_{err}$ is given by
\begin{align}
&p_{err}=\mathbb{P}(\hat{\tilde{C}}_{N}(x)<\eta/2|\tilde{C}(x)\geq \eta-\epsilon)\mathbb{P}(\tilde{C}(x)\geq \eta-\epsilon)\notag\\
&+\mathbb{P}(\hat{\tilde{C}}_{N}(x)\geq \eta/2|\tilde{C}(x)\leq \epsilon)\mathbb{P}(\tilde{C}(x)\leq \epsilon)\label{eq:errprob}\\
&\leq \mathbb{P}(\hat{\tilde{C}}_{N}(x)<\eta/2|\tilde{C}(x)=\eta-\epsilon)\mathbb{P}(\tilde{C}(x)\geq \eta-\epsilon)\notag\\
&+\mathbb{P}(\hat{\tilde{C}}_{N}(x)\geq \eta/2|\tilde{C}(x)=\epsilon)\mathbb{P}(\tilde{C}(x)\leq \epsilon).\notag
\end{align}
Lets consider the term $\mathbb{P}(\hat{\tilde{C}}_{N}(x)\geq \eta/2|\tilde{C}(x)=\epsilon)$, then conditioned on $\tilde{C}(x)=\epsilon$ we get
\begin{align*}
&\mathbb{P}(\hat{\tilde{C}}_{N}(x)\geq \eta/2)=\mathbb{P}(\hat{\tilde{C}}_{N}(x)-E[\hat{\tilde{C}}_{N}(x)]\geq \eta/2-\epsilon)\\
&=\mathbb{P}(\sum_{l=1}^{N}S_l -E[\sum_{l=1}^{N}S_l]\geq \frac{N}{2}(\eta/2-\epsilon)),
\end{align*}
since $\hat{\tilde{C}}_{N}(x)$ is unbiased estimator of $\tilde{C}(x)$ and so $E[\hat{\tilde{C}}_{N}(x)]=\tilde{C}(x)$.
Applying Hoeffding's inequality leads to
\begin{align*}
&\mathbb{P}(\hat{\tilde{C}}_{N}(x)\geq \eta/2|\tilde{C}(x)=\epsilon)\leq e^{-\frac{2(\frac{N}{2}(\eta/2-\epsilon))^2}{4N(A(\overrightarrow{r}))^2}}=e^{-\frac{(\eta/2-\epsilon)^2 N}{8(A(\overrightarrow{r}))^2}}.
\end{align*}
Taking $N=N_s$ given in Eqn. (\ref{eq:samples}) it follows that
\begin{align}
&\mathbb{P}(\hat{\tilde{C}}_{N}(x)\geq \eta/2|\tilde{C}(x))\leq \nu.
\end{align}
Following similar analysis one can show that $\mathbb{P}(\hat{\tilde{C}}_{N}(x)<\eta/2|\tilde{C}(x)=\eta-\epsilon)\leq\nu$. Thus
\begin{align*}
&p_{err} \leq \nu\left(P(\tilde{C}(x)\geq \eta-\epsilon)+P(\tilde{C}(x)\leq \epsilon)\right)\leq \nu,
\end{align*}
as desired.

\bibliographystyle{IEEETran}
\bibliography{references}

@article{spe,
  title={Randomized quantum algorithm for statistical phase estimation},
  author={Wan, Kianna and Berta, Mario and Campbell, Earl T},
  journal={Physical Review Letters},
  volume={129},
  number={3},
  pages={030503},
  year={2022},
  publisher={APS}
}

@article{lin2022heisenberg,
  title={Heisenberg-limited ground-state energy estimation for early fault-tolerant quantum computers},
  author={Lin, Lin and Tong, Yu},
  journal={PRX quantum},
  volume={3},
  number={1},
  pages={010318},
  year={2022},
  publisher={APS}
}

@article{o2019quantum,
  title={Quantum phase estimation of multiple eigenvalues for small-scale (noisy) experiments},
  author={O’Brien, Thomas E and Tarasinski, Brian and Terhal, Barbara M},
  journal={New Journal of Physics},
  volume={21},
  number={2},
  pages={023022},
  year={2019},
  publisher={IOP Publishing}
}

@article{dutkiewicz2021heisenberg,
  title={Heisenberg-limited quantum phase estimation of multiple eigenvalues with a single control qubit},
  author={Dutkiewicz, Alicja and Terhal, Barbara M and O’Brien, Thomas E},
  journal={arXiv preprint arXiv:2107.04605},
  year={2021}
}

@article{poulin2014trotter,
  title={The Trotter step size required for accurate quantum simulation of quantum chemistry},
  author={Poulin, David and Hastings, Matthew B and Wecker, Dave and Wiebe, Nathan and Doherty, Andrew C and Troyer, Matthias},
  journal={arXiv preprint arXiv:1406.4920},
  year={2014}
}

@article{babbush2015chemical,
  title={Chemical basis of Trotter-Suzuki errors in quantum chemistry simulation},
  author={Babbush, Ryan and McClean, Jarrod and Wecker, Dave and Aspuru-Guzik, Al{\'a}n and Wiebe, Nathan},
  journal={Physical Review A},
  volume={91},
  number={2},
  pages={022311},
  year={2015},
  publisher={APS}
}

@article{babbush2019quantum,
  title={Quantum simulation of chemistry with sublinear scaling in basis size},
  author={Babbush, Ryan and Berry, Dominic W and McClean, Jarrod R and Neven, Hartmut},
  journal={npj Quantum Information},
  volume={5},
  number={1},
  pages={92},
  year={2019},
  publisher={Nature Publishing Group UK London}
}

@article{sarkar2024scalable,
  title={Scalable quantum circuits for exponential of Pauli strings and Hamiltonian simulations},
  author={Sarkar, Rohit Sarma and Chakraborty, Sabyasachi and Adhikari, Bibhas},
  journal={arXiv preprint arXiv:2405.13605},
  year={2024}
}

@book{nielsen2010quantum,
  title={Quantum computation and quantum information},
  author={Nielsen, Michael A and Chuang, Isaac L},
  year={2010},
  publisher={Cambridge university press}
}

@article{kitaev1995quantum,
  title={Quantum measurements and the Abelian stabilizer problem},
  author={Kitaev, A Yu},
  journal={arXiv preprint quant-ph/9511026},
  year={1995}
}

@article{lavielle2005using,
  title={Using penalized contrasts for the change-point problem},
  author={Lavielle, Marc},
  journal={Signal processing},
  volume={85},
  number={8},
  pages={1501--1510},
  year={2005},
  publisher={Elsevier}
}

@article{aminikhanghahi2017survey,
  title={A survey of methods for time series change point detection},
  author={Aminikhanghahi, Samaneh and Cook, Diane J},
  journal={Knowledge and information systems},
  volume={51},
  number={2},
  pages={339--367},
  year={2017},
  publisher={Springer}
}

@article{truong2020selective,
  title={Selective review of offline change point detection methods},
  author={Truong, Charles and Oudre, Laurent and Vayatis, Nicolas},
  journal={Signal processing},
  volume={167},
  pages={107299},
  year={2020},
  publisher={Elsevier}
}

@article{berry2019qubitization,
  title={Qubitization of arbitrary basis quantum chemistry leveraging sparsity and low rank factorization},
  author={Berry, Dominic W and Gidney, Craig and Motta, Mario and McClean, Jarrod R and Babbush, Ryan},
  journal={Quantum},
  volume={3},
  pages={208},
  year={2019},
  publisher={Verein zur F{\"o}rderung des Open Access Publizierens in den Quantenwissenschaften}
}

@article{killick2012optimal,
  title={Optimal detection of changepoints with a linear computational cost},
  author={Killick, Rebecca and Fearnhead, Paul and Eckley, Idris A},
  journal={Journal of the American Statistical Association},
  volume={107},
  number={500},
  pages={1590--1598},
  year={2012},
  publisher={Taylor \& Francis}
}

@misc{pyscf,
      title={The Python-based Simulations of Chemistry Framework (PySCF)}, 
      author={Qiming Sun and Timothy C. Berkelbach and Nick S. Blunt and George H. Booth and Sheng Guo and Zhendong Li and Junzi Liu and James McClain and Elvira R. Sayfutyarova and Sandeep Sharma and Sebastian Wouters and Garnet Kin-Lic Chan},
      year={2017},
      eprint={1701.08223},
      archivePrefix={arXiv},
      primaryClass={physics.chem-ph},
      url={https://arxiv.org/abs/1701.08223}, 
}

@misc{openfermion,
      title={OpenFermion: The Electronic Structure Package for Quantum Computers}, 
      author={Jarrod R. McClean and Kevin J. Sung and Ian D. Kivlichan and Yudong Cao and Chengyu Dai and E. Schuyler Fried and Craig Gidney and Brendan Gimby and Pranav Gokhale and Thomas Häner and Tarini Hardikar and Vojtěch Havlíček and Oscar Higgott and Cupjin Huang and Josh Izaac and Zhang Jiang and Xinle Liu and Sam McArdle and Matthew Neeley and Thomas O'Brien and Bryan O'Gorman and Isil Ozfidan and Maxwell D. Radin and Jhonathan Romero and Nicholas Rubin and Nicolas P. D. Sawaya and Kanav Setia and Sukin Sim and Damian S. Steiger and Mark Steudtner and Qiming Sun and Wei Sun and Daochen Wang and Fang Zhang and Ryan Babbush},
      year={2019},
      eprint={1710.07629},
      archivePrefix={arXiv},
      primaryClass={quant-ph},
      url={https://arxiv.org/abs/1710.07629}, 
}

@article{jordanwigner,
  title = {{\"U}ber das Paulische {\"A}quivalenzverbot},
  author = {Jordan, P. and Wigner, E.},
  journal = {Zeitschrift f{\"u}r Physik},
  volume = {47},
  number = {9-10},
  pages = {631--651},
  year = {1928},
  publisher = {Springer},
  doi = {10.1007/BF01331938}
}

\end{document}